\documentclass[%
 reprint,
superscriptaddress,
 amsmath,amssymb,
 aps,
 prl,
`floatfix,
]{revtex4-1}

\usepackage{graphicx}
\usepackage{dcolumn}
\usepackage{bm}
\usepackage{hyperref}

\usepackage{pdfpages}

\begin{document}

\preprint{APS/123-QED}

\title{Coupled networks and networks with bimodal frequency distributions are equivalent}

\author{Bastian Pietras} 
\affiliation{%
 MOVE Research Institute Amsterdam \& Institute for Brain and Behavior Amsterdam, Faculty of Behavioural and Movement Sciences, Vrije Universiteit Amsterdam, van der
Boechorststraat 9, Amsterdam 1081 BT, The Netherlands}

\author{Nicol\'as Deschle}%
\affiliation{%
 MOVE Research Institute Amsterdam \& Institute for Brain and Behavior Amsterdam, Faculty of Behavioural and Movement Sciences, Vrije Universiteit Amsterdam, van der
Boechorststraat 9, Amsterdam 1081 BT, The Netherlands}

\author{Andreas Daffertshofer}
\email{a.daffertshofer@vu.nl}
\affiliation{%
 MOVE Research Institute Amsterdam \& Institute for Brain and Behavior Amsterdam, Faculty of Behavioural and Movement Sciences, Vrije Universiteit Amsterdam, van der
Boechorststraat 9, Amsterdam 1081 BT, The Netherlands}




\date{\today}

\begin{abstract}
Populations of oscillators can display a variety of synchronization patterns depending on the oscillators' intrinsic coupling and the coupling between them. We consider two coupled, symmetric (sub)populations with unimodal frequency distributions and show that  the resulting synchronization patterns may resemble those of a single population with bimodally distributed frequencies. Our proof of the equivalence of their stability, dynamics, and bifurcations, is based on an Ott-Antonsen ansatz. The generalization to networks consisting of multiple (sub)populations vis-\`a-vis networks with multimodal frequency distributions, however, appears impossible.
\begin{description}
\item[PACS numbers]
\pacs{05.45.Xt}05.45.Xt 
\end{description}
\end{abstract}

\pacs{Valid PACS appear here}
\maketitle

The Kuramoto model is seminal for describing synchronization patterns in networks of phase oscillators. 
It has been investigated to great detail in numerous studies using different approaches; see, e.g., \cite{Acebron_Review2005,Rodrigues2016} for review.
The analytical treatment typically relies on the formation of a common variable, the so-called order parameter, and seeks to pinpoint its dynamics.
The more recently suggested ansatz by Ott and Antonsen \cite{OttAntonsen2008} proved particularly fruitful for analyzing this dynamics.
It applies to the thermodynamic limit, i.e.
to infinitely large populations, and it contains major simplifications including the 'parametrization' of the phase distribution's Fourier transform.
Abrams and co-workers \cite{AbramsSolv2008} were the first to describe the dynamics of two coupled populations using the Ott-Antonsen ansatz, confirming earlier results based on perturbation techniques \cite{MontbrioTwoPop2004,BarretoNetworks2008}; see also Laing's extension including heterogeneity and phase lags \cite{Laing2009}.
Similarly, Kawamura and co-workers
\cite{Kawamura2010} derived a collective phase sensitivity function to describe synchronization across subpopulations, but they assumed only very weak coupling between them. 
A detailed bifurcation analysis of these dynamics without such restrictions, however, is still missing.

We  discuss a network of two populations of Kuramoto oscillators with unimodally distributed natural frequencies. 
The dynamics will be compared with that of a single population of oscillators with bimodally distributed frequencies.
The latter case has been extensively studied by Martens and co-workers
\cite{MartensExactResults2009}.
Here we prove that a two-population network does fully resemble the case of one network with bimodally distributed frequencies. By contrast, the extension to more than two populations or to multimodal frequency distributions remains a challenge, if at all possible.

Let us consider two symmetrical populations.
Both consist of $N$ phase oscillators $\theta_{\sigma,k}$, with $\sigma = 1,2$ and $k = 1, \dots, N$. The oscillators have natural frequencies $\omega_{\sigma,k}$ distributed according to a Lorentzian of width $\Lambda_1\!=\!\Lambda_2\!=\!\Lambda$ that are centered around $+\varpi_0$ and $-\varpi_0$, respectively.
We assume all-to-all coupling within each population with strength $K_\mathrm{int}$, and also all-to-all coupling across populations with strength $K_\mathrm{ext}$.
The corresponding dynamics obeys the form
\begin{eqnarray}
\label{Kuramoto-equations}
\dot{\theta}_{\sigma,k}
&=& \omega_{\sigma,k} + \frac{K_\mathrm{int}}{N} \sum_{j=1}^N \sin(\theta_{\sigma,j} - \theta_{\sigma,k})  +
\nonumber \\ 
&& \hspace*{6em}+ \frac{K_\mathrm{ext}}{N} \sum_{j=1}^N \sin(\theta_{{\sigma'},j} - \theta_{\sigma,k})
\end{eqnarray}
with $(\sigma,{\sigma'}) = (1,2)$ or $(2,1)$.
We consider the limit $N \to \infty$ and introduce continuous, time-dependent distribution functions $f_\sigma$ of the subpopulations. The integral of  $f_\sigma$ over phase and frequency defines the (local) order parameters 
\begin{align*}
 z_{\sigma} = \int_\mathbb{R}\int_0^{2 \pi}  f_{\sigma} (\omega,\theta,t) \, \mathrm{e}^{i\theta} \,d\theta\, d\omega \ ,
\end{align*}
i.e. a (circular) 'mean value' for each population $\sigma$.
The Ott-Antonsen ansatz \cite{OttAntonsen2008} incorporates the $2\pi$-periodicity of $f_\sigma$ and further simplifies its Fourier series to a single Fourier component $\alpha_\sigma(\omega,t)$, i.e.
\begin{align*}
f_{\sigma}(\omega, \theta,  t) = \frac{g_{\sigma}(\omega)}{2\pi} \left\lbrace 1 \!+\! \left[ \sum_{n=1}^{\infty} \alpha_{\sigma}(\omega, t)^n \mathrm{e}^{i n \theta} + \mathrm{c.c.}  \right] \right\rbrace .
\end{align*}
With the normalization
\begin{align*}
\int_0^{2\pi} f_{\sigma} (\omega,\theta,t) \, d\theta = g_{\sigma}(\omega) := \frac{\Lambda}{\pi} \frac{1}{\left(\omega - \omega_{\sigma}\right)^2 + \Lambda^2} \ ,
\end{align*}
where $\omega_{1/2}=\pm\varpi_0$, the dynamics of the order parameters $z_\sigma$ reduces to 
\begin{eqnarray}
\label{z_dynamics}
\dot{z}_{\sigma} &=& - \left(\Lambda \mp i \varpi_0\right) z_{\sigma} + \frac{K_\mathrm{int}}{2} z_{\sigma} \left(1-\left|z_{\sigma}\right|^2\right) + 
\nonumber \\
&& \hspace*{10em}
+\frac{K_\mathrm{ext}}{2} \left(z_{{\sigma'}} - z_{\sigma}^2 z_{{\sigma'}}^{\ast}\right) .
\end{eqnarray}
Since $g_\sigma(\omega)$ are continuous, non-constant frequency distributions, the Ott-Antonsen manifold comprises the entire dynamics  \cite{OttAntonsen2009}.
Next, we rewrite the order parameters as $z_{\sigma}= \rho_{\sigma} \mathrm{e}^{i\phi_{\sigma}}$ such that 
with the assumed symmetry $\rho := \rho_1 = \rho_2$ the system \eqref{z_dynamics} transforms into
\begin{align}
\label{reduced_eqn}
\begin{split}
\dot{\rho} & = - \Lambda \rho + \frac{\rho}{2} (1-\rho^2) \left[K_\mathrm{int} + K_\mathrm{ext} \cos\psi\right] \\
\dot{\psi} & = 2\varpi_0 - K_\mathrm{ext} (1 + \rho^2) \sin\psi \ ;
\end{split}
\end{align}
here we introduced the mean relative phase between the subpopulations as $\psi\! =\!\phi_2\!-\!\phi_1$.
Finally, we rescale the parameters by means of $\tau\!=\!K_\mathrm{int}\!\cdot\!{t}$, $\kappa\!=\!K_\mathrm{ext}/K_\mathrm{int}$, $\Delta\!=\!2\Lambda/K_\mathrm{int}$ and $\omega_0\!=\!2\varpi_0/K_\mathrm{int}$, substitute $q \!=\!\rho^2$, and transform $q(t) \to q(\tau)$ as well as $\psi(t) \to \psi(\tau)$ if not stated otherwise \footnote{We consider $K_\mathrm{int} \neq 0$ and note that the scaling does not affect the quality of bifurcations, i.e. the original and scaled systems are topologically equivalent.}.
Then, we find for $0<\rho< 1$ 
\begin{align}
\label{scaled_eqn}
\begin{split}
\dot q & = q \left[ 1 - \Delta -q + \kappa(1-q) \cos\psi \right]\\
\dot \psi & = \omega_0 - \kappa(1+q)\sin\psi \ ;
\end{split}
\end{align}
from hereon the dot notation refers to the derivative with respect to $\tau$.
The system \eqref{scaled_eqn} resembles Eqs. (25\&26) in \cite{MartensExactResults2009} with the addition of the bifurcation parameter $\kappa$. For $\kappa\!=\!1$ both systems agree entirely \footnote{Our unscaled system \eqref{reduced_eqn} is an exact representation of Eqs. (22\&23) for $\tilde{K}=K/2$ in the notation used in \cite{MartensExactResults2009}.}.
As we will show, however, the additional bifurcation parameter $\kappa$, does not alter the qualitative bifurcation scheme of our network.
Hence, we can understand the bimodal formulation as an equivalent representation of the network consisting of two symmetric subpopulations.

\begin{figure}[hbt]
\includegraphics[width= .8\columnwidth]{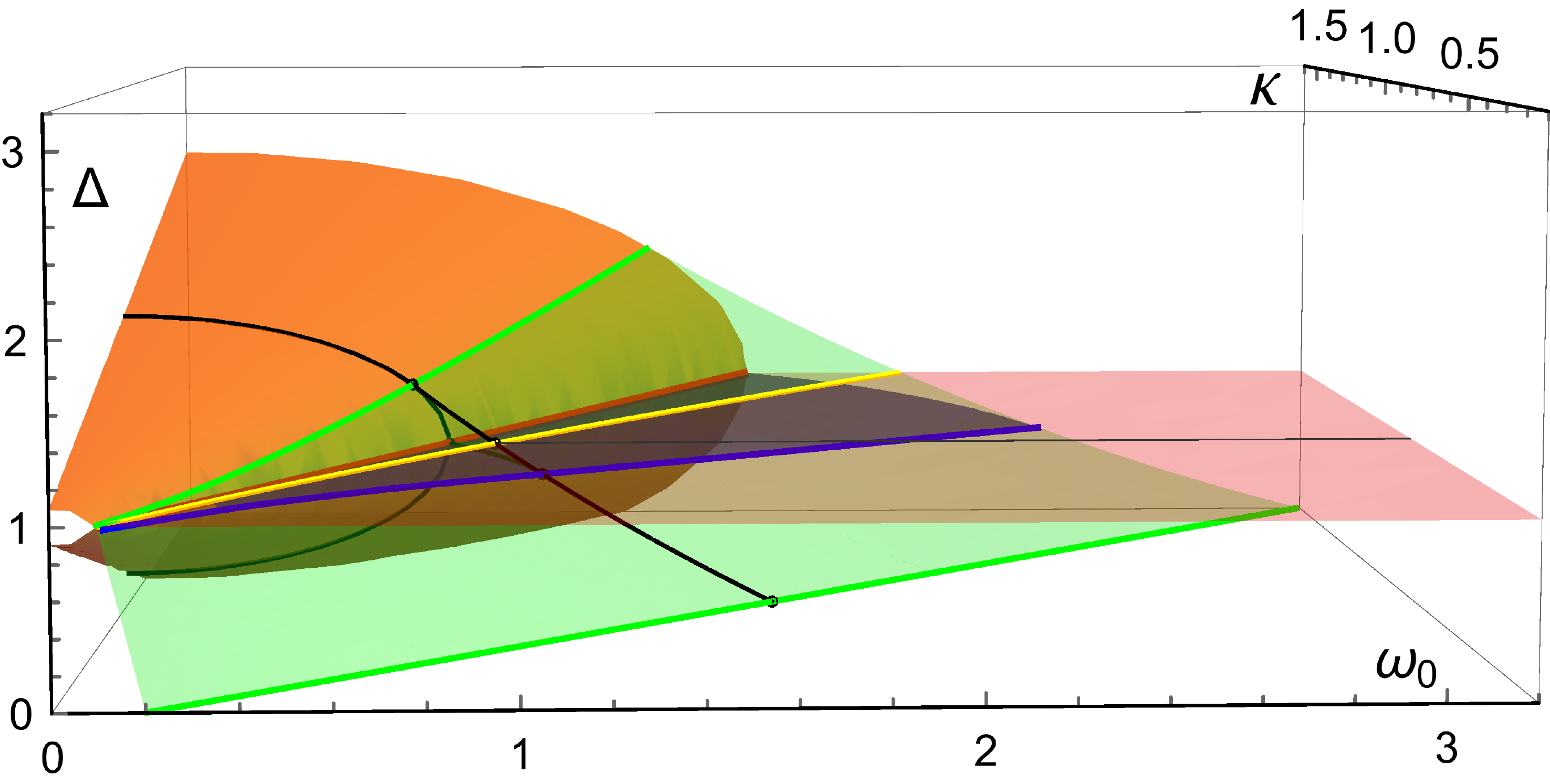} 
\caption{Bifurcation boundaries.
Red plane: Hopf, orange cone: transcritical, green plane (within green lines): saddle node, blue: homoclinic bifurcation.
Blue line: Bogdanov-Takens curve, yellow: intersection of Hopf and SN, black lines: cross-section at $\kappa=0.8$, see also Fig. \ref{bistab_region}.}
\label{bifurcation_boundaries}
\end{figure}

Before discussing \eqref{scaled_eqn} in more detail, we briefly analyze the stability of the fully incoherent state $q\!=\!0$. 
Following Martens et al.
\cite{MartensExactResults2009}, we linearize \eqref{z_dynamics} around $z_1=z_2=0$ and find two pairs of degenerated eigenvalues
\begin{equation}
\label{zero_ev}
\lambda_{1/3} = \lambda_{2/4} = 1 - \Delta \mp \sqrt{\kappa^2 - \omega_0^2}
\end{equation}
expressed in the aforementioned, rescaled parameters.
Given the rotational invariance of the incoherent state, we expected this degeneracy.
The incoherent state is linearly stable if and only if the real parts of these eigenvalues are less than or equal to zero.
Using $\kappa\geq0$ and $\omega_0\geq 0$ we find the stability boundary as
\begin{gather}
\Delta = 1+
\left\{
\begin{array}{ll}\displaystyle
\sqrt{\kappa^2 - \omega_0^2} &\text{for } \ \kappa \geq \omega_0\\ \displaystyle
0 &\text{otherwise}
\end{array} \ ,
\right.\ 
\end{gather}
which can be confirmed by perturbing the uniform distribution $f(\omega,\theta,t) = (2\pi)^{-1}$; see Montbri\'o and co-workers
\cite{MontbrioTwoPop2004} or Okuda and Kuramoto \cite{OkudaKuramoto1991}.
Crossing this boundary for $\kappa\!\geq\!\omega_0$ corresponds to a degenerated transcritical bifurcation, while crossing the half line $\Delta\!=\!1$ resembles a degenerated supercritical Hopf bifurcation; see Fig. \ref{bifurcation_boundaries}, where the red plane displays the Hopf bifurcation and the orange cone the transcritical one.

Coming back to the system \eqref{scaled_eqn} we realize that its fixed points satisfy
$1\!-\!\Delta\!-\!q\!=\!\kappa(1-q)\cos \psi$ and $\omega_0\!=\!\kappa(1\!+\!q)\sin\psi$.
Combining these using 
$\cos^2\psi\!+\!\sin^2\psi\!=\!1$ yields
$\kappa^2 = \left(\left(1\!-\!\Delta\!-\!q\right)\!/\!\left(1\!-\!q\right)\right)^2 + \left(\omega_0/\!\left(1\!+\!q\right)\right)^2$,
or, equivalently,
\begin{equation}
\label{hypersurface}
\omega_0 = \pm \frac{1+q}{1-q} \sqrt{\Delta(2-2q-\Delta) - (1-\kappa^2)(1-q)^2} 
\end{equation}
as the implicit form of a hyperplane of fixed points $q_s=q_s(\omega_0,\Delta,\kappa)$.
After inserting $\partial \omega_0/\partial q\!=\!0$ in \eqref{hypersurface}, the solution $\omega_0= \omega_0(\Delta,\kappa)$ forms a surface (green in Fig. \ref{bifurcation_boundaries}) across which a saddle-node bifurcation appears.
If both subpopulations contain oscillators with identical natural frequencies $\omega_{\sigma}$,  i.e. if $\Delta\!=\!0$, then the saddle-node curve emerges from $\kappa=\omega_0/2$. 
We stress this because in the literature the saddle-node curve has only been approximated numerically, while here we find that the Ott-Antonsen ansatz allows for deriving an analytical solution in a straightforward manner.
The saddle-node plane starts at $(\omega_0,\Delta)=(2\kappa,0)$ and approaches tangentially the transcritical bifurcation plane at 
$(\omega_0,\Delta) =1/4\!\left(\sqrt{8\kappa^2\!-\!2\!+\!2\sqrt{1+8\kappa^2}},3\!+\!\sqrt{1\!+\!8\kappa^2}\right)$.
This solution is consistent with the intersection point $(\omega_0,\Delta)_{\kappa=1}=\left(\sqrt{3}/2,3/2\right)$ reported in \cite{MartensExactResults2009}.

\begin{figure}[htbp]
\begin{minipage}{0.48\columnwidth}
\includegraphics[width= .9\textwidth]{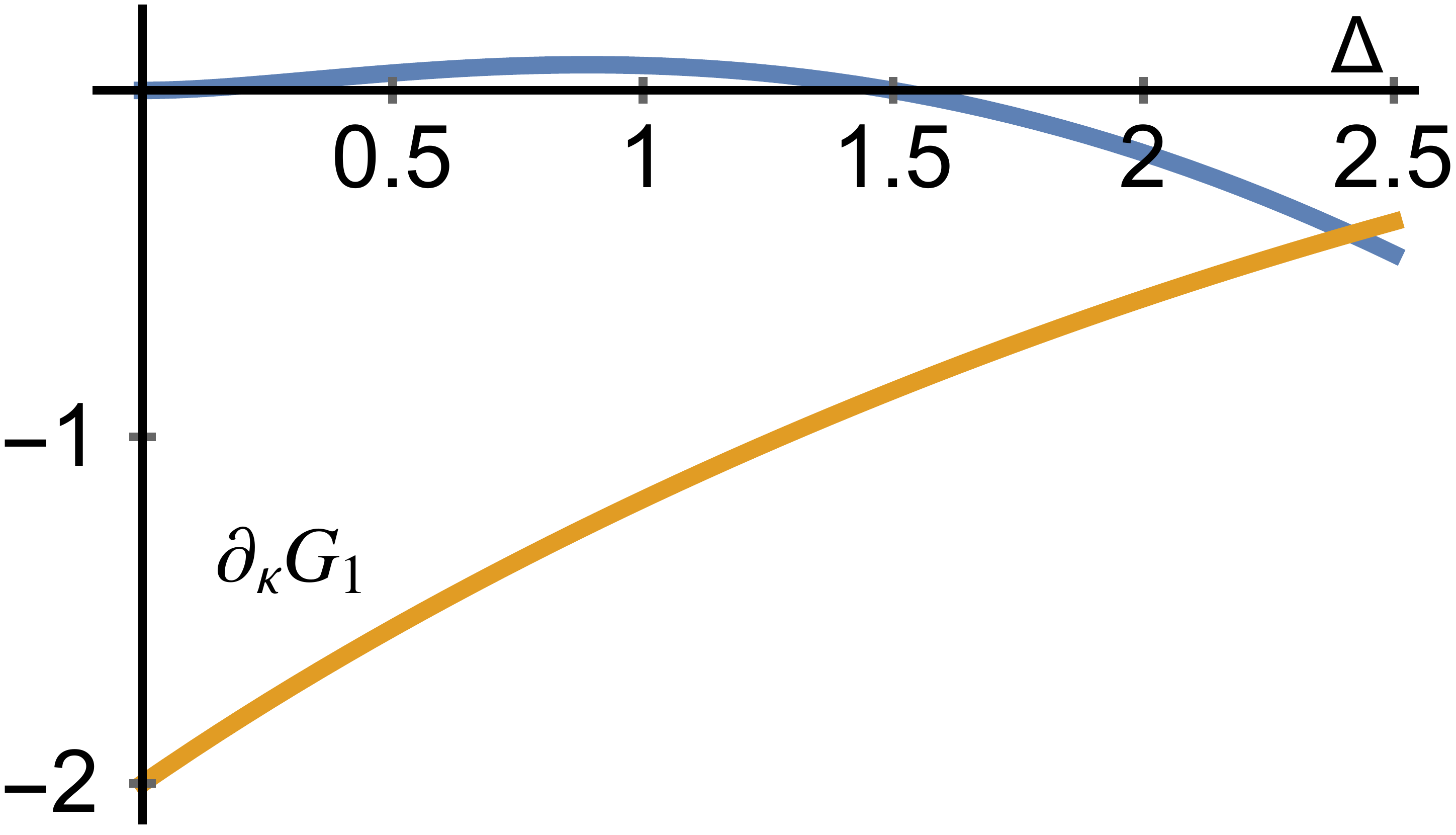}
\end{minipage}
\hfill
\begin{minipage}{0.48\columnwidth}
\includegraphics[width= .9\textwidth]{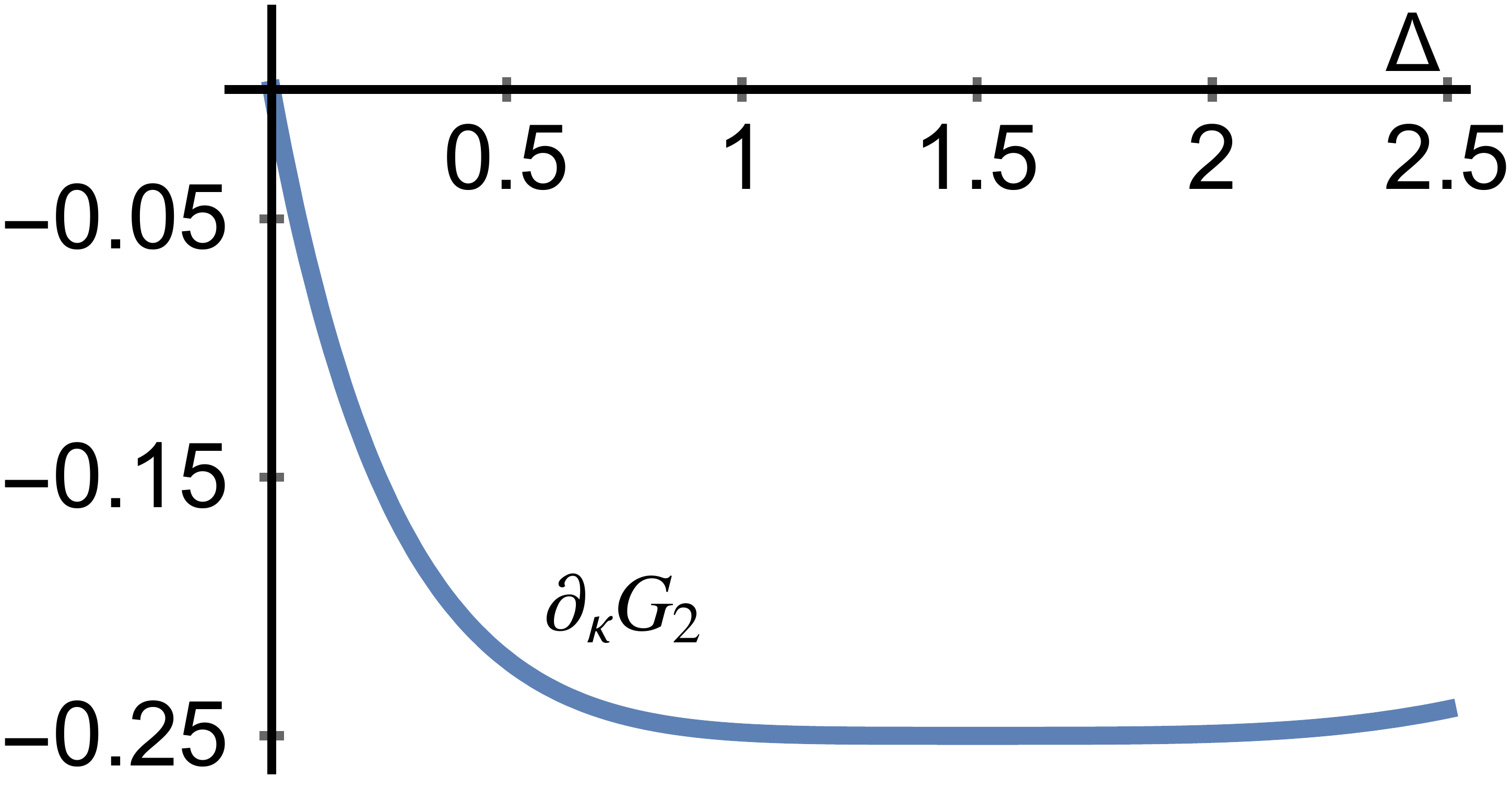}
\end{minipage}
\caption{Partial derivatives of $\partial_\kappa G$ along the saddle-node-plane at $\kappa = 1$;
left panel: $\partial_\kappa G_1(\Delta)$, right panel: $\partial_\kappa G_2(\Delta)$.}
\label{delcG}
\end{figure}

Can a change in $\kappa$ lead to new bifurcation behavior? To show that this is not the case, let $G_1(q,\psi;\Delta,\omega_0,\kappa)$ denote the right-hand side of \eqref{scaled_eqn} and define $G_2(q,\psi;\Delta,\omega_0,\kappa)\!=\!\det\!\left\{\partial_{(q,\psi)}G_1(q,\psi;\Delta,\omega_0,\kappa)\right\}$.
For $\kappa = 1$ it follows that
\begin{equation}
\boldsymbol{G}(q,\psi;\Delta,\omega_0,\kappa) := \binom{G_1(q,\psi;\Delta,\omega_0,\kappa)}{G_2(q,\psi;\Delta,\omega_0,\kappa)}= 0
\end{equation}
along the saddle-node curve; cf. Eq. (33) in \cite{MartensExactResults2009}.
According to the implicit function theorem, there is no qualitative change in the $(\Delta,\omega_0)$-bifurcation diagram if
\begin{equation}
\partial_\kappa \boldsymbol{G}(q,\psi;\Delta,\omega_0,\kappa) \neq 0
\label{partialG}
\end{equation}
for any neutrally stable fixed point $(q,\psi;\Delta,\omega_0,\kappa)\!=:\!\boldsymbol{x}$.
Here, however, we have to extend this to a family of fixed points $\boldsymbol{x}_s = \boldsymbol{x}(\Delta)$ along the saddle-node curve parametrized by $\Delta$.
Therefore, if \eqref{partialG} holds for a fixed point $\boldsymbol{x}_1$, i.e. if $\partial_\kappa \boldsymbol{G}(\boldsymbol{x}_1) \neq 0$, then we still may end up at another point $\boldsymbol{x}_2$ on that curve.
We circumvent this case by also requiring for any arbitrary $a \in \mathbb{R}$
\begin{equation}
\partial_\kappa G_1(q,\psi;\Delta,\omega_0,\kappa) \, \neq \, a\!\cdot\!\partial_{\Delta} G_1(q,\psi;\Delta,\omega_0,\kappa)
\end{equation}
at every point along the saddle-node curve.
Fig.  \ref{delcG} shows that the inequality \eqref{partialG} holds for all $\boldsymbol{x}_s$.
We note that, because $\dot{\psi}$ is independent of $\Delta$, it suffices to consider only the second equation of $\partial_\kappa G_1$, which is non-zero for $0\!\leq\!\Delta\!<\!4$.
That is, the bifurcation diagram is persistent against (small) perturbations around $\kappa\!= \!1$
and there are no bifurcations of co-dimension larger than $2$. 

For co-dimension $2$, Martens and co-workers suggested the existence of Bogdanov-Takens points on the saddle-node plane below the Hopf bifurcation that can be identified numerically.
In fact, the reduced dynamics \eqref{scaled_eqn} has a Jacobian along the saddle-node plane that is (conjugate to) a diagonal matrix with only one zero eigenvalue in the parameter range under study. 
This underlines the saddle-node character of that plane, but more importantly, it shows that these equations cannot be exploited for bifurcation points of co-dimension $2$.

\begin{figure}[htbp]
\centering{
\includegraphics[width= .8\columnwidth]{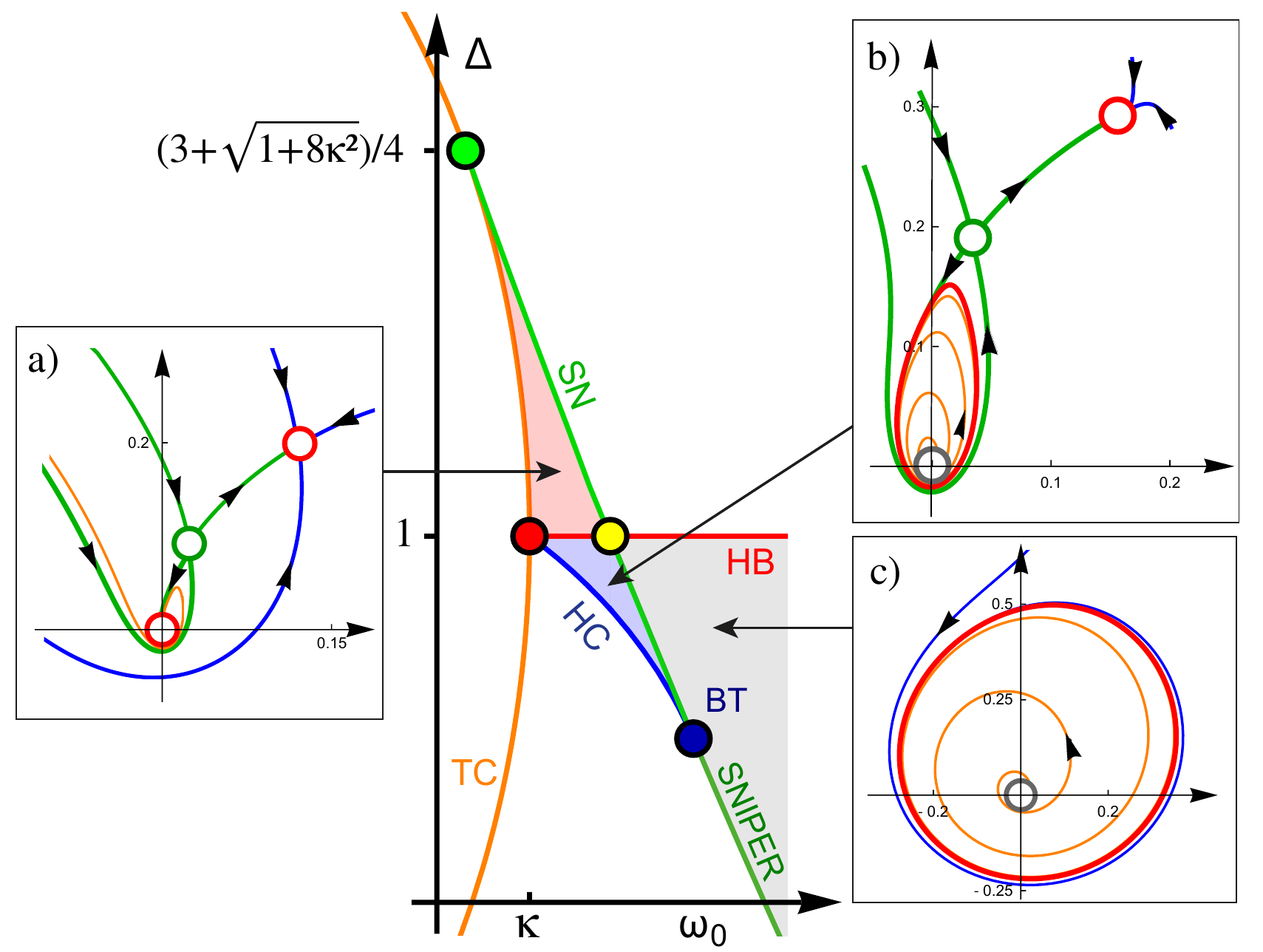}}
\caption{
Bifurcation boundaries: cross-section of Fig. \ref{bifurcation_boundaries} at $\kappa < 1$;
red: Hopf, orange: transcritical, green: saddle node, blue: homoclinic, blue point: Bogdanov-Takens bifurcation.
Insets: $(q,\psi)$-phase portraits (in polar coordinates) in their specific parameter regions, red circle: stable fixed point, gray: unstable fixed point, green: saddle point.
The bistability region (red/blue) overlaps with the oscillatory regime (blue/gray). (a) Coexistence of two stable fixed points, (b) a stable fixed point outside a stable limit cycle, (c) the more regular, stable limit cycle away from the SN curve.
}
\label{bistab_region}
\end{figure}

In the {\em Supplementary Material} we provide more details of the bifurcation scheme including numerical simulations. 
The latter demonstrate the existence of a multistability region; cf. Fig. \ref{bistab_region} and Martens et al.'s Figs.~5 \& 7a. Multistability has been reported independently in \cite{MontbrioTwoPop2004,MartensExactResults2009,OkudaKuramoto1991,Laing2009}.
The red parameter region, bounded by the transcritical cone (orange curve), the Hopf plane (red) and the saddle-node plane (green), reveals the coexistence of another stable, but non-trivial fixed point next to the stable incoherent solution (separated by a saddle point).
In the blue parameter region left to the saddle-node plane and below the red Hopf plane, the incoherent solution has undergone a supercritical Hopf bifurcation such that a stable limit cycle coexists with the pair of stable fixed and saddle points.
For the stability properties of our solutions  we refer to section IV. in \cite{MartensExactResults2009}. Due to the  equivalence of both the bimodal and the two subpopulation system, the stability results there can be readily adopted. Note that the equivalence also holds when introducing a time delay; see {\em Supplementary Material}. 

Particularly interesting for future applications are the limit cycle oscillations in the $(q\cos\psi, q\sin\psi)$-plane shown in  Figs. \ref{bistab_region}(b) and (c).
There, both $q(t+T) = q(t)$ and $\psi(t+T) = \psi(t) \mod 2\pi$ hold for all $t\in \mathbb{R}$ given a fixed period length $T=T(\Delta,\omega_0,K)$.
We study these oscillations in more detail by introducing the global complex-valued order parameter $z = (z_1\!+\!z_2)/2$, whose magnitude $|z| = R$ reads
\footnote{The absolute value of the global order parameter $z$ reads in general: $R = |z| 
= \frac{1}{2} \left| \rho_1 \mathrm{e}^{i\phi_1} + \rho_2 \mathrm{e}^{i\phi_2}  \right|  
= \frac{1}{2} \sqrt{\rho_1^2+\rho_2^2 +2\rho_1\rho_2 \cos(\phi_2\!-\!\phi_1)}$.
}
\begin{align}
R = \frac{\rho}{\sqrt{2}}\sqrt{1+\cos\psi}  
\end{align}
with $\rho = \sqrt{q}$.
If $\dot{\psi}(t) \neq 0$, then $R(t)$ will oscillate.
We would like to note that in this case oscillations in $R$ would be even observable without $q$ being periodic. However, for all parameter values outside the oscillatory regime, the dynamics contains stable fixed points at which obviously $\dot{\psi} = 0$, i.e. $R\to\text{const}$. As can be seen in Fig. \ref{bistab_region}(b), the limit cycle is deformed: it is neither circular nor symmetric about the origin. Then, also $q$ oscillates, i.e. not only the global order parameter $R$ oscillates, but so do the local ones $\rho=\rho_1=\rho_2$.
For larger $\omega_0$ the limit cycle gains symmetry, but does not become a perfect circle. Hence oscillations contain higher harmonics; see Fig. \ref{bistab_region}(c).
Future studies will address more details of the parameter-dependency of frequency and amplitude of the $\rho$ and $R$ oscillations as well as their phase difference.

Given that two coupled networks and networks with bimodal frequency distributions are equivalent, it appears obvious to seek for generalizations, here the case of more than two subpopulations.
Anderson and co-workers studied communities of oscillators in systems with multiple subpopulations \cite{AndersonOA2012}.
They included mixes of attractive and repulsive couplings (in our notation $K_\mathrm{int}$ and $K_\mathrm{ext}$ should differ in sign) rendering the dynamics too diverse for analytical treatment.
Closer to our approach, however, is the work by Komarov and Pikovsky \cite{KomarovPikovsky2011} who showed a variety of synchronization characteristics as well as the emergence of chaotic states in the case of three positively coupled subpopulations.

\begin{figure}[hbt]
\hspace*{-0.3cm}
\includegraphics[width= .9\columnwidth]{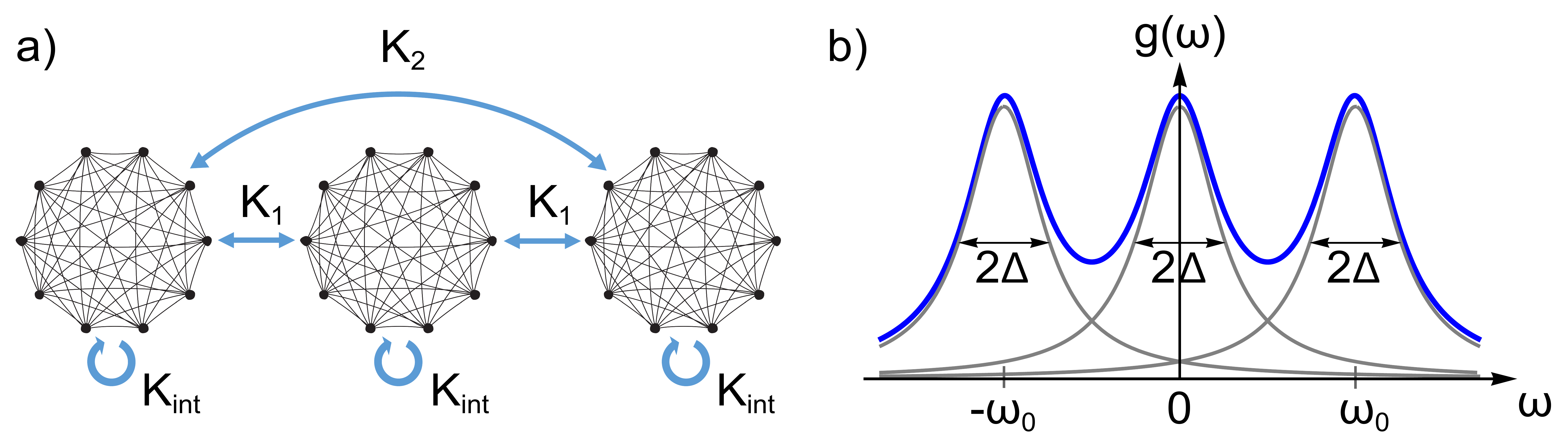}
\caption{(a)  Three all-to-all coupled networks; (b) symmetric trimodal frequency distribution function.}
\label{tripop}
\end{figure}

We sketch the case of
three subpopulations with unimodal frequency distribution each: $g_\sigma(\omega) =  (\Lambda/\pi)/((\omega\!-\!(-\varpi_0,0,+\varpi_0))^2\!+\!\Lambda^2)$ with peaks at ($-\varpi_0,0,+\varpi_0$) \footnote{$\varpi_0$ is assumed to be sufficiently large to guarantee isolated peaks and all distributions have width $\Lambda$.}. This is compared to oscillators with a symmetric, trimodal frequency distribution: $g(\omega) = \beta\!\cdot\!g_1(\omega)\!+\!\alpha\!\cdot\!g_2(\omega)\!+\!\beta\!\cdot\!g_3(\omega)$
with $\alpha= (4\varpi_0^2\!-\!2 \Lambda^2)/(12\varpi_0^2)$, and $\beta = (4\varpi_0^2\!+\! \Lambda^2)/(12 \varpi_0^2)$.
The systems read
\begin{subequations}
\begin{align}
\dot{\theta}_{\sigma,k} &= \omega_{\sigma,k} + \sum_{\tau = 1}^3 \frac{K_{\sigma,\tau}}{N} \sum_{j=1}^N \sin(\theta_{\tau,j}-\theta_{\sigma,k}) 
\label{three_subpop}\\
\dot{\theta}_k &= \omega_k + \frac{K}{3N} \sum_{j=1}^{3N} \sin(\theta_j - \theta_k) \ , 
\label{trimodal}
\end{align}
\end{subequations}
where $K_{\sigma,\tau} = K_{|\sigma-\tau|}$ with $K_0$ denoting the internal coupling strength $K_{\mathrm{int}}$ within each population, $K_1$ the coupling strength between adjacent populations, and $K_2$ that between distant populations, see Fig. \ref{tripop}.
In \eqref{trimodal} we have $k = 1,\dots, 3N$. 
Again, we introduce (local) order parameters $z_\sigma = \rho_\sigma e^{i\phi_{\sigma}}$.
Since the two outer populations are considered symmetric, we use $\rho_{13} \equiv \rho_1 = \rho_3$ and that $\phi_2 - \phi_1 = \phi_2 - \phi_3 :=\psi$.
By this we find the dynamics of \eqref{trimodal} after rescaling $\tau=(K/2)\cdot{t}$ and $\omega_0=2\varpi_0/K$ and 
$\Delta=2\Lambda/K$ and $\kappa_\alpha=\alpha$ and $\kappa_\beta=\beta$ as
\begin{align*} 
\dot{\rho}_{13} &= \rho_{13}\left[ -\Delta\!+\!\left(1\!-\!\rho_{13}^2\right)\!\left(\kappa_\alpha \frac{\rho_2}{\rho_{13}} \cos\psi\!+\!\kappa_\beta\!\left(1\!+\!\cos2\psi\right)\right) \right] \nonumber\\
\dot{\rho}_2 &= \rho_2 \left[ -\Delta\!+\!\left(1\!-\!\rho_2^2\right)\!\left( \kappa_\alpha\!+\!2\kappa_\beta \frac{\rho_{13}}{\rho_2} \cos\psi\right) \right]\\
\dot{\psi} &= \omega_0\!-\!\left(1\!+\!\rho_{13}^2\right)\!\left(\kappa_\alpha \frac{\rho_2}{\rho_{13}}\sin\psi\!+\!\kappa_\beta\sin2\psi  \right) \ .\nonumber
\end{align*}
Accordingly we rescale the system \eqref{three_subpop} using $K\!=\!K_{\mathrm{int}}\!+\!K_1\!+\!K_2$
  and $\tau = (K/2)\cdot{t}$, $\Delta = 2 \Lambda/K$, $\omega_0 = 2 \varpi_0 / K$ and abbreviate $\kappa_{\alpha,\beta} = 2 K_{1,2}/K$, which yields
\begin{align*}
\dot{\rho}_{13} &= \rho_{13} \left[-\Delta\!+\!\left(1 \!-\!\rho_{13}^2\right)\!\left(\kappa_0\!+\!\kappa_\alpha \frac{\rho_2}{\rho_{13}} \cos\psi\!+\!\kappa_\beta \cos2\psi\right)\right] \nonumber\\
\dot{\rho}_2 &= \rho_2 \left[ - \Delta\!+\! \left(1\!-\! \rho_2^2\right)\!\left(\kappa_0\!+\!2\kappa_\alpha\frac{\rho_{13}}{\rho_{2}}  \cos\psi\right) \right]\\
\dot{\psi} &= \omega_0\!-\!\left(1\!+\!\rho_{13}^2\right)\!\left[ \kappa_\alpha \frac{\rho_2}{\rho_{13} }\sin\psi \!+\!\kappa_\beta \sin2\psi \right] \ , \nonumber
\end{align*} 
where $\kappa_0 = 1-\kappa_\alpha - \kappa_\beta$.
Both systems can display a richer dynamical behavior than the dynamics \eqref{scaled_eqn} since they, e.g., contain coupling terms of first and second harmonics, that may result in a $2\!:\!1$ phase synchronization.
When it comes to linking the two, we realize that they are only identical for the special case
\begin{align*}
\kappa_\alpha = \kappa_\beta = \frac{1}{3} \quad \Rightarrow \quad  \alpha = \beta \ .
\end{align*}
As $\alpha$ and $\beta$ only differ by $\Lambda^2/(4\varpi_0)$, this implies $\Lambda \to 0$, hence the distribution function will consist of three $\delta$-peaks and the inhomogeneity is strongly reduced.
As a consequence, the Ott-Antonsen manifold does not exhibit the whole dynamics of our system \cite{OttAntonsen2009} and our description will remain incomplete,
as has been found by Martens in \cite{Martens2010} for even stronger symmetry assumptions in a network of three populations.

We conclude that, while stability, dynamics, and bifurcations of a symmetric two population system of phase oscillators are equivalent to a single population with bimodal frequency distribution, one cannot readily generalize this to a multimodal/multiple subpopulation system.

\subsection{Acknowledgements}
This project has received funding form the European Union's Horizon 2020 research and innovation program under the Marie Sk{\l}odowska-Curie grant agreement \#642563 (COSMOS).

%

~\newpage\includepdf[pages=1,scale=1,pagecommand={\thispagestyle{empty}}]{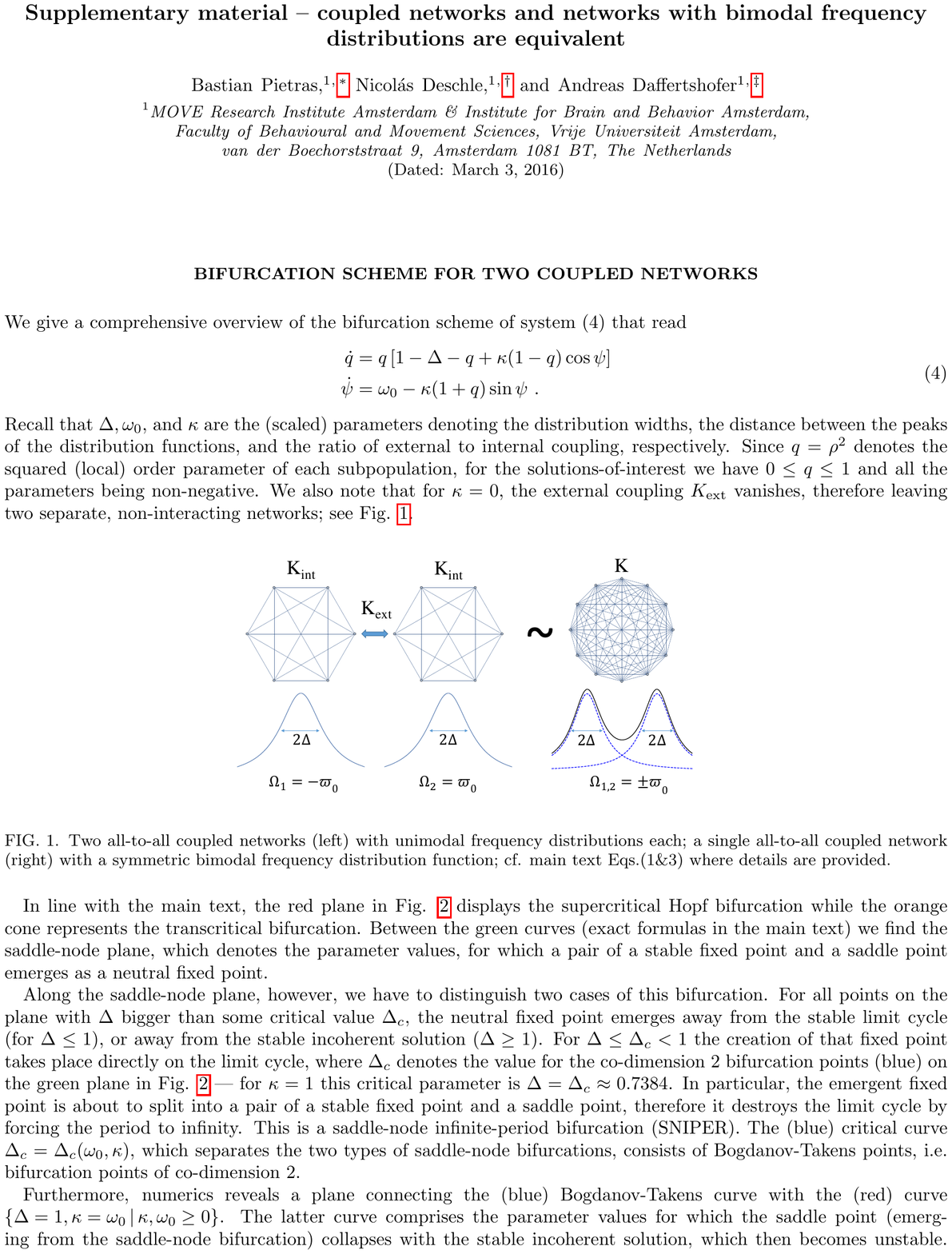}
~\newpage\includepdf[pages=2,scale=1,pagecommand={\thispagestyle{empty}}]{Pietras_etal_TwoNetworks_Supplement.pdf}
~\newpage\includepdf[pages=3,scale=1,pagecommand={\thispagestyle{empty}}]{Pietras_etal_TwoNetworks_Supplement.pdf}

\end{document}